\documentstyle[twoside,fleqn,espcrc2,epsfig]{article}
\begin{document}

\newcommand{\ber}{\begin{eqnarray}}
\newcommand{\eer}{\end{eqnarray}}
\newcommand{\spz}{\hspace{0.7cm}}
\newcommand{\br}{\langle}
\newcommand{\kt}{\rangle}
\newcommand{\um}{\frac12}
\newcommand{\beq}{\begin{equation}}
\newcommand{\eeq}{\end{equation}}
\newcommand{\sg}{\sigma_p}
\newcommand{\sm}{\sigma_c}
\newcommand{\se}{\sigma_e}
\newcommand{\lf}{\langle L_f \rangle}
\newcommand{\llf}{\left\langle | L_f | \right\rangle}
\newcommand{\la}{\langle L_a \rangle}
\newcommand{\tf}{{\rm Tr}_f}
\newcommand{\nt}{N_\tau}
\newcommand{\ns}{N_\sigma}
\newcommand{\f}{\beta_f}
\newcommand{\ba}{\beta_a}
\newcommand{\bv}{\beta_v}
\newcommand{\bt}{\beta}
\newcommand{\bvc}{\beta_{vc}}
\newcommand{\lm}{\lambda}
\newcommand{\gm}{\gamma}

\hyphenation{ }

\title{Impact of $Z_2$ monopoles and vortices on the deconfinement transition}

\author{Saumen Datta\address{Mehta Research Institute, Chhatnag Road,
Jhusi,Allahabad 211019, India}
and R. V. Gavai\address{Department of Theoretical Physics, Tata Institute of 
Fundamental Research, Homi Bhabha Road, Mumbai 400005, India} }

\begin{abstract}
Suppressing $Z_2$-monopoles shifts the line of deconfinement transitions in the
coupling plane  of the $SU(2)$ lattice gauge theory with a mixed Villain 
form of action but it still continues to behave also like the bulk transition 
line.  Separate deconfinement and bulk phase transitions are found on the same 
lattice, suggesting the two to be indeed coincident at higher adjoint couplings.
Universality is restored when both monopoles and vortices are suppressed. 

\end{abstract}

\maketitle
\vskip-0.5cm
\section{UNIVERSALITY}

Since most numerical lattice field theory investigations are necessarily 
carried out for a finite value of the lattice cut-off $a$,  it seems 
imperative that universality of the results so obtained is verified by 
employing other forms of lattice actions.  A study of the universality of 
the deconfinement transition for SU(2) gauge theory, which has already been
extensively investigated for the Wilson action\cite{wilson},  
was made in Ref. \cite{gavai}, for the Bhanot-Creutz action \cite{bhanot}
\beq
S = \sum_p \left[ \f \left( 1 - {{\rm Tr}_f U_p \over 2} \right)
 + \ba  \left( 1 - {{\rm Tr}_a U_p\over 3}  \right) \right]
\label{eq.bc}\eeq
Here the summation runs over all the plaquettes of the lattice and
the subscript $a$($f$) indicates that the trace is taken in the adjoint 
(fundamental) representation.  Later, a Villain form of action\cite{caneschi}, 
defined by
\beq
S = \sum_p \left[ \f +\bv - {{ \left( \f + \bv \sg \right) 
{\rm Tr}_f U_p } \over 2} \right]
\label{eq.action}\eeq
was also used\cite{stephenson} for similar studies with essentially similar 
results, where $\sg$ are auxiliary $Z_2$-variables defined on the plaquettes 
and the partition function has an additional sum over all possible values of 
the $\sg$ variables as well.  The Wilson action corresponds to setting 
$\ba$ or $\bv$ to zero above.  Simulations\cite{gavai,stephenson} on $ \ns^3 
\times \nt $ lattices with $\nt$ = 4 showed surprising results for the 
above actions.  While the second order 
deconfinement transition point for the Wilson action entered the coupling plane
as a line of second order transitions, the transition turned 
first order for large enough $\ba$ or $\bv$.  The order parameter for the
deconfinement transition acquired a nonzero value discontinuously there
and the exponent of the corresponding susceptibility changed
from the Ising model value of about 1.97 to 3. If the change of the
order of the deconfinement transition were to persist at a finite $\ba$
with increasing temporal lattice size $\nt$, i.e., in the continuum
limit, it would be a serious violation of universality. On the other
hand, the line of deconfinement transitions was found to coincide with the 
known bulk transition lines\cite{bhanot,caneschi}.  Additional studies
with varying $\nt$ further revealed that the line scarcely moves 
in the region where a strong first order deconfinement transition is
observed\cite{manu}.  

Inspired by the results\cite{i2} for the $SO(3)$ lattice gauge
theory, we investigated the finite temperature phase diagram of the
mixed action (\ref{eq.action}) with suppression of the $Z_2$-monopoles 
and vortices by addition of chemical potentials for them.  These terms
are irrelevant in the naive continuum limit.  

\vskip-0.5cm
\section{MONOPOLES AND VORTICES}
\label{sc.nomono}

The $Z_2$-monopoles are suppressed\cite{i2} by the addition of a 
chemical potential term, $\lm \sum_c \left( 1 - \sm \right)$, to the mixed 
Villain action (\ref{eq.action}).  The summation runs over all the 
elementary 3-cubes of the lattice, and $\sm = \prod_{p \in \partial c} \sg $. 
Note that in the classical continuum limit one still obtains the same 
$\lm$-independent continuum relation, $4 g^{-2} = \f + \bv$. Following 
the $SO(3)$ results, we took $\lm$ = 1 for our simulations in the 
entire $\f$ - $\bv$ plane. Our each iteration consisted of heatbath sweeps
for all the gauge links, followed by those for the $Z_2$- variables. 
A fraction of the links (arbitrarily chosen to be $\frac{1}{4}$) were 
$Z_2$-rotated subject to a probability determined by the $\f$ term at
the end of each iteration to reduce the otherwise enormous
autocorrelations for large $\lm$ simulations.  Measurements 
were made after every iteration. Using hysteresis runs of 15000 
iterations per point we mapped out the phase diagram on an $8^3\times4$ 
lattice.  

\vskip-1.0cm
\begin{figure}[htbp]\begin{center}
\epsfig{height=6cm,width=7.5cm,file=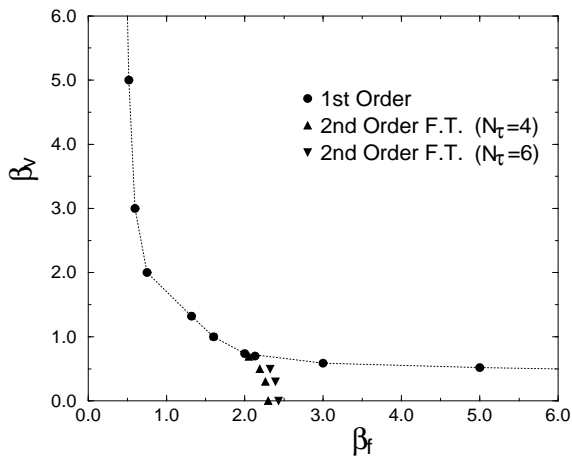}
\vskip-1.0cm
\caption{The phase diagram for the action (\protect\ref{eq.action}) with
monopole suppression on an $8^3 \times 4$ lattice. }
\vskip-1.0cm
\label{fg.bulkmono}\end{center}\end{figure}

For $\bv > \f$,  first order transitions, with discontinuities in both  
the average plaquette $P=\langle \frac{1}{2} {\rm Tr}_f U_p \rangle$ and 
the fundamental Polyakov loop $\llf$, were observed. These transition points 
are shown by filled circles in Fig. \ref{fg.bulkmono}. Since $\llf$
becomes nonzero discontinuously at these couplings,  it clearly indicates a 
first order deconfinement phase transition. A second order deconfinement phase 
transition coincident with a first order bulk phase transition, signaled by 
the discontinuity in the average plaquette at the same location, is also 
possible. This general behavior is 
very similar to what was observed for $\lm$ =0 \cite{stephenson}. In
contrast to that case, however, one now also has a first order phase 
transition for $ \bv < \f$, as shown in Fig.  \ref{fg.bulkmono}. 
Here the observable $P_a = \langle \frac{1}{2} \sg . \tf U_p \rangle$ 
displays a sizeable discontinuity.  A qualitatively new feature of the 
phase diagram in Fig. \ref{fg.bulkmono} thus is the absence of any end point 
for the transition line because of  the new line of transitions 
coming from the large $\f$ side along which the deconfinement order parameter
is nonzero on both sides of the transition.

As a continuation of the deconfinement transition on the Wilson axis, 
we looked for a deconfinement transition at $\bv$ = 0.3, 0.5 and 0.7.  
The transition point was located approximately from the sharp but continuous 
rise of $\llf$. From the peak heights of the $|L_f|$ - susceptibility, 
$\chi_{|L_f|}$, on $\ns^3 \times 4$ lattices for $\ns$ = 8, 12 and 16, 
its critical exponent was obtained at each $\bv$.  A linear fit to ${\rm ln} ~
(\chi_{|L_f|})_{max} = \omega ~{\rm ln} ~ \ns$ gave $\omega = 1.91\pm0.02,
1.87\pm0.05$ and $1.92\pm0.05$ for $\bv$ = 0.7, 0.5 and 0.3,
respectively. They indicate second order transitions
and are in agreement with the $\bv$ = 0 exponent and the Ising
model exponent. The average plaquette $\langle P \rangle$ from these
runs was smooth everywhere, and the corresponding susceptibility
peaks did not sharpen with $\ns$  at all, indicating a lack of bulk
transition at these points. These transition points, shown in Fig.
\ref{fg.bulkmono} by triangles, are therefore pure finite temperature
transitions. Similar results for $\nt$=6 are also shown.

\vskip-0.5cm
\begin{figure}[htbp]\begin{center}
\epsfig{height=5.5cm,width=7.5cm,file=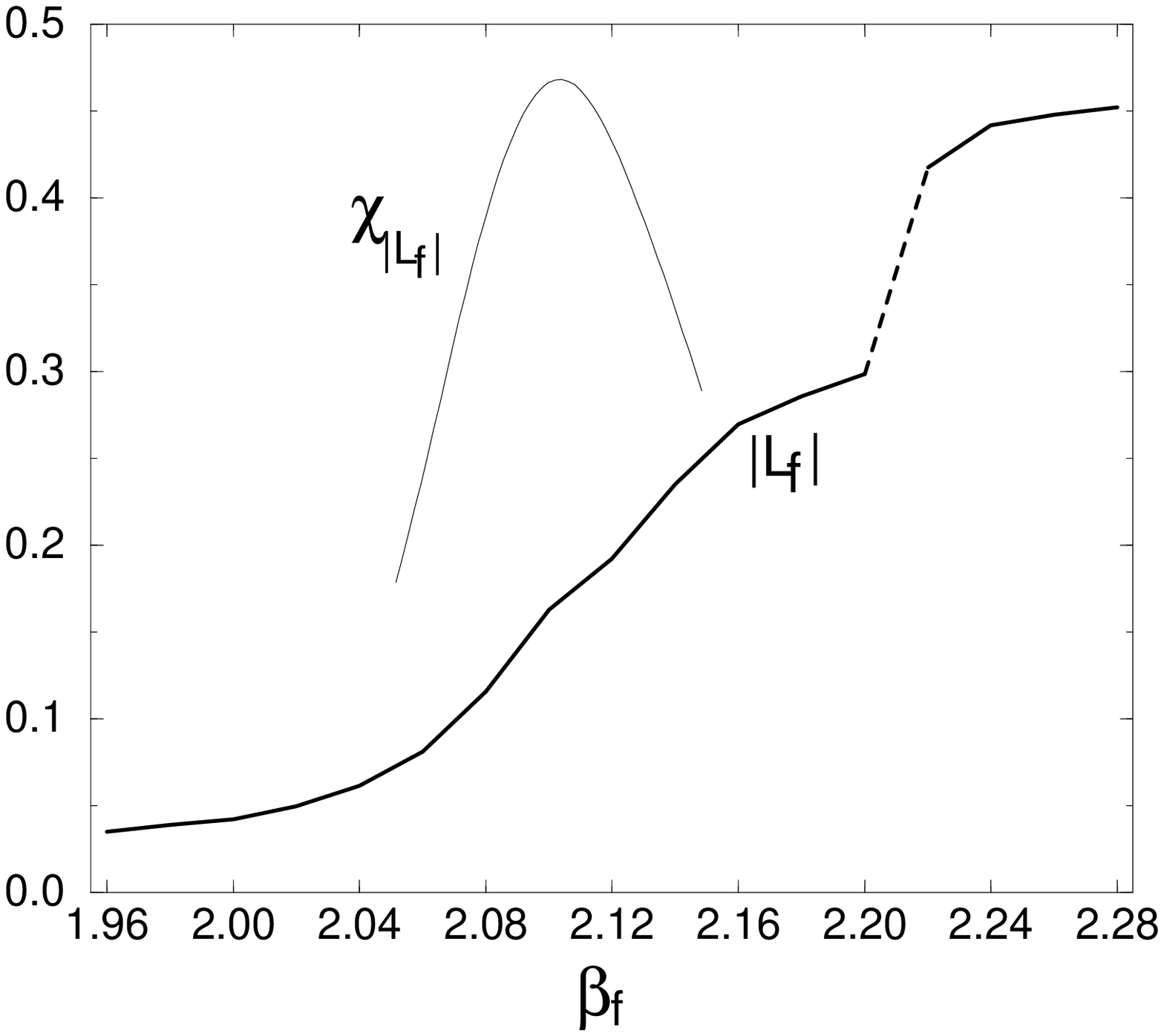}
\vskip-1.0cm
\caption{ $\llf$ and its susceptibility as a function of $\f$ at $\bv$ = 0.7. }
\vskip-1.0cm
\label{fg.twotr}\end{center}\end{figure} 

The $\lm$ = 1 simulations for the
mixed Villain action thus lend a strong credibility to the hypothesis
that the deconfinement transition line for possibly a large range of
$\nt$ merges with the bulk transition line.  Since the latter branches
out in this case and exhibits no end point, the merger is easy to
observe numerically: the small $\bv$ region has only a deconfinement
transition line and the large $\f$ region has only a bulk transition line,
while they seem to be coincident for $\bv \ge 0.7$.

\vskip-0.5cm
\begin{figure}[htbp]\begin{center}
\epsfig{height=5cm,width=7.5cm,file=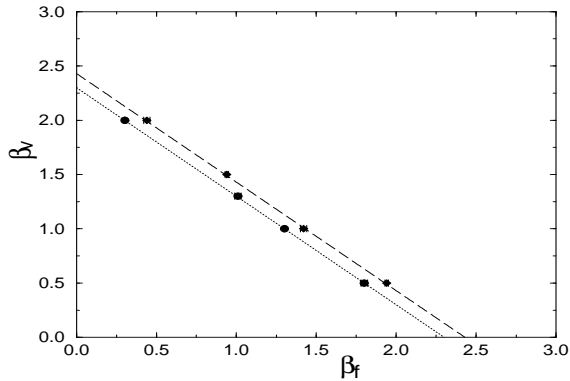}
\vskip-1.0cm
\caption{The deconfinement transition points in the ($\f$, $\bv$) 
plane for $\nt$ = 4 (circles) and 6 (diamonds) lattices.}
\vskip-1.0cm
\label{fg.dcpln}\end{center}\end{figure}

A possible litmus test of the coincidence scenario is to see two separate
transitions on the same lattice.  Fig. \ref{fg.twotr}
shows the results of such a test at $\bv$ = 0.7 on $8^3\times4$ lattices.
It shows the susceptibility, $\chi_{|L_f|}$, obtained from the longer run 
mentioned above along with the order parameter $\llf$ obtained from
a hysteresis run from a hot start.  It clearly shows a second order 
deconfinement transition taking place first at $\f \sim 2.1$, followed by 
a bulk phase transition later at $\f \sim 2.2$.

In order to suppress the $Z_2$-electric vortices in addition to the magnetic
monopoles, we added to the action (\ref{eq.action}) another irrelevant
term, $ \gm \sum_l \left( 1 - \se \right)$, where $\se = \prod_{p \in 
\hat \partial l} \sg $.  For sufficiently large $\gm$, one expects that the 
bulk transition line in Fig. \ref{fg.bulkmono}, caused presumably by the 
condensation of electric loops, will also be suppressed.  
As $\lm \to \infty$, the monopole term is frozen and the plaquette variables 
$\sg$ are replaced by products over corresponding $Z_2$-link variables. 

Following the same procedure as above, and using a heat-bath
algorithm for both the gauge and $Z_2$-variables, we studied
the phase diagram on $\nt=4$ and 6 lattices for $(\lm,\gm)=(\infty,5.)$.
Fig. \ref{fg.dcpln} shows the only transition points found, which are 
Ising-like second order deconfinement transitions and obey 
$\f + \bv \approx \beta_c^W$, where $\beta_c^W$ is the deconfinement 
transition point for the Wilson action. The transition lines are
also consistent with the expected continuum limit behavior of this action.

\vskip-0.5cm
\section{CONCLUSIONS}
\label{sc.summary}

Our numerical simulations for the monopole-suppressed action showed an 
interesting phase diagram which was different from that of the original theory.
Nevertheless, it too had the paradoxical coincidence of bulk and
deconfinement transitions. The bulk transition line in 
this case had no end point and  the change of the order 
of the deconfinement phase transition occurred as the two lines merged. 
We showed the presence of two phase transitions on the 
same finite lattice in the vicinity of the point of merger (See Fig.
\ref{fg.twotr}). 

A further suppression of the $Z_2$ electric vortices got
rid of the bulk transitions completely and yielded only lines of second order 
deconfinement transitions, in agreement with universality.
Since the terms added to the action in the process do not contribute in the 
naive continuum limit, one can formally attribute the anomalous behavior of 
the deconfinement transition lines for both the mixed actions 
to the presence of bulk transitions.

\end{document}